\documentclass{PoS}
\usepackage{amsmath}
\usepackage{booktabs}
\graphicspath{{figs/}}
\renewcommand{\O}{\mathcal{O}}

\title{$B_c$ spectroscopy using highly improved staggered quarks}

\ShortTitle{$B_c$ spectroscopy using HISQ}

\author{\speaker{Andrew Lytle} \\
        INFN, Sezione di Roma Tor Vergata, Via della Ricerca Scientifica 1,
        00133 Roma RM, Italy\\
        E-mail: \email{andrew.lytle@roma2.infn.it}}

\author{Brian Colquhoun \\
         High Energy Accelerator Research Organization (KEK),
         Tsukuba 305-0801, Japan\\}

\author{Christine Davies\\
        SUPA, School of Physics and Astronomy, University of Glasgow, 
        Glasgow, G12 8QQ, UK\\}

\author{Jonna Koponen \\
        INFN, Sezione di Roma Tor Vergata, Via della Ricerca Scientifica 1,
        00133 Roma RM, Italy\\}

\abstract{We report on a calculation of
$B_c$ ground state and radial excitation energies, 
obtained from heavy-charm 
highly improved staggered quark (HISQ) correlators computed on MILC
gauge ensembles, with lattice spacings down to $a=0.044$ fm. 
Using HISQ valence quarks on progressively finer lattices allows
us to simulate up to the $b$-quark mass.
In particular we focus on the $B_c(2S)$ energy, which we compare with
$\O(\alpha_s)$-improved non-relativistic QCD results computed on the same 
ensembles and recent experimental results from ATLAS.}

\FullConference{The 36th Annual International Symposium on Lattice Field Theory - LATTICE2018\\
		22-28 July, 2018\\
		Michigan State University, East Lansing, Michigan, USA.}

\begin{document}
\section{Introduction}
Recent years have seen a great deal of experimental progress in the study of
$B_c$ mesons at the LHC, including precise 
lifetime measurements~\cite{Aaij:2014bva,Aaij:2014gka,Sirunyan:2017nbv}, 
observation of new
hadronic~\cite{Aaij:2013gia,Anderlini:2014dha,Aaij:2014asa}, 
and semileptonic~\cite{Aaij:2017tyk} decay channels,
and excited states~\cite{Aad:2014laa,Aaij:2017lpg}.
Precision lattice QCD calculations of the $B_c$ system
can provide a number of opportunities when
combined with new experimental measurements, including
new avenues for determination of $|V_{cb}|$~\cite{Wingate:2017unz}, 
shedding light on flavour anomalies~\cite{Colquhoun:2016osw,Cohen:2018dgz}, 
and identification of resonance peaks in data.

Based on LHC Run 1 data the ATLAS collaboration observed a resonant
structure at the 5$\sigma$ level in decays to the $B_c$ ground state,
which they identified as a radial $B_c$ excitation
with energy 6842$(4)_{\text{syst}} (5)_{\text{stat}}$ MeV~\cite{Aad:2014laa}.
This state was subsequently searched for by the LHCb 
collaboration,
however no corresponding structure was observed in their Run 1 data, despite 
having a higher yield of $B_c$ signal candidates~\cite{Aaij:2017lpg}.
Therefore the location of this state remains an open question.

Here we report our progress on a calculation to determine the $B_c(2S)$ energy
directly from lattice QCD data, using the heavy HISQ methodology 
described in more detail below, as well as using lattice NRQCD. 
Section~\ref{details} explains
the details of the calculation, while Section~\ref{results}
gives the status of results for the $B_c$ ground state
and $B_c$ $2S$-$1S$ splitting.

\section{Details of calculation} \label{details}
We work on ensembles of $n_f=2+1+1$ gauge configurations generated by the
MILC collaboration~\cite{Bazavov:2010ru}. All of the ensembles used here have unphysically
heavy pion masses in the sea, while the sea strange and charm quark masses
are near their physical values. We use the highly improved staggered quark
(HISQ) action~\cite{Follana:2006rc} to compute charm and heavy valence quarks.
Throughout the calculation we use charm
valence quarks with $m^{\text{val}}_c$ near to its physical value. 
We can check the effect of any mistuning by varying $m^{\text{val}}_c$.

Because $a m_b$ is large, even on our finest lattice spacing ensemble, 
quantities calculated at $m_b$ directly would have potentially large
discretisation artifacts proportional to powers of $(am_b)^2$, making
controlled continuum extrapolations unfeasible. To get around this we
work with heavy quark masses (generically referred to as $m_h$), and 
compute the quantity of interest keeping $a m_h < 0.8$, and using a range
of $m_h$ values on multiple lattice spacings.
In this way we can fit the $(am_h)^{2n}$ discretisation effects and recover the
physical dependence on $m_h$, which we model as a power series in
$1/M_{\eta_h}$. Finally we evaluate this function at $M_{\eta_b}$ 
to determine physical results.
The ensembles and parameter values used in this calculation are collected in
Table~\ref{tab:ens}.

We tie together charm and heavy quark propagators to construct 
zero momentum two-point functions $C(t)$, with the quantum numbers
to create and destroy Goldstone pseudoscalar mesons. 
$C(t)$ is a sum of exponentials,
\begin{equation} \label{exp}
C(t) = \sum_i a_i e^{-M_i t} + (-1)^t b_i e^{-\tilde{M}_i t} + (t \to T-t) \,.
\end{equation}
Here we are interested in the
lowest and first excited state energies $M_1$ and $M_2$ 
corresponding to the $H_c$ ground state and first radial excitation.
We use multi-exponential Bayesian fits to determine the amplitudes and energies
from Eq.~\eqref{exp}.

In addition to the fully relativistic calculation we also calculate
the $B_c$ $2S$-$1S$ splitting using $\O(\alpha_s)$-improved 
non-relativistic QCD (NRQCD) for the $b$-quark on the c-5, f-5, and sf-5
ensembles. 
The NRQCD propagators are obtained using the time evolution operator
constructed from the NRQCD Hamiltonian. 
The NRQCD $b$-quark propagator is combined
with a HISQ charm propagator on each gauge background, 
and averaged 
to construct the two-point functions. 
More details of the procedure can be found in~\cite{Colquhoun:2015oha}.

\begin{table} \label{tab:ens}
\centering
\begin{tabular}{llllllll}
\toprule
  ens & $\beta$ &   $am_l$ &  $am_s$ & $am_c$ & $N_{s}\times{N_t}$ & $am_c^{\text{val}}$ &    $am_h^{\text{val}}$ \\
\midrule
  c-5 &    6.00 &   0.0102 &  0.0509 &  0.635 &       24$\times$64 &               0.635 &                 \quad- \\
  f-5 &    6.30 &   0.0074 &   0.037 &  0.440 &       32$\times$96 &               0.434 &              0.6,\,0.8 \\
 f-10 &    6.30 &   0.0036 &   0.036 &  0.430 &       48$\times$96 &               0.439 &              0.6,\,0.8 \\
 sf-5 &    6.72 &   0.0048 &   0.024 &  0.286 &      48$\times$144 &               0.274 &  0.4,\,0.5,\,0.6,\,0.8 \\
 uf-5 &    7.00 &  0.00316 &  0.0158 &  0.188 &      64$\times$192 &               0.188 &              0.6,\,0.8 \\
   \, &      \, &       \, &      \, &     \, &                 \, &               0.195 &  0.4,\,0.5,\,0.7,\,0.9 \\
\bottomrule
\end{tabular}

\caption{Summary of ensembles and run parameters used in this work.
The ensemble specifications in the left-hand column appear in the
figure legends, and correspond to ``coarse'' ($a \approx 0.12$ fm),
``fine'' ($a \approx 0.09$ fm),
``superfine'' ($a \approx 0.06$ fm) and ``ultrafine'' ($a \approx 0.044$ fm)
lattice spacings and have $m_s/m_l = 5$, or 10 in the case of the `f-10'
ensemble.
}
\end{table}

\section{Results} \label{results}
\subsection{$B_c$ mass}
\begin{figure}
\includegraphics[width=\textwidth]{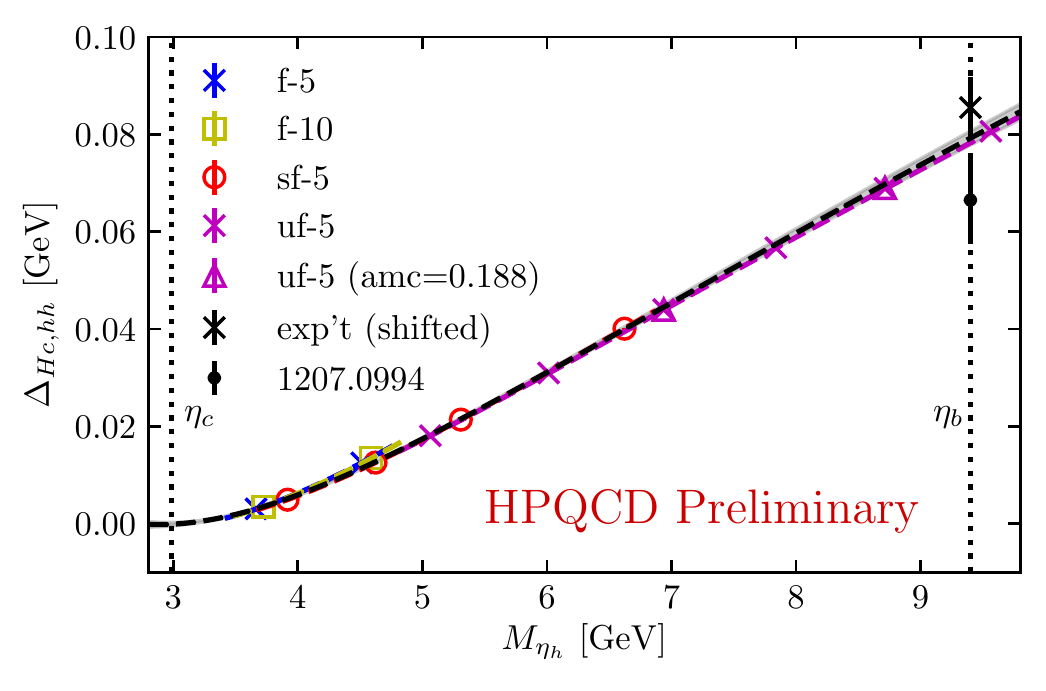}
\caption{The quantity $M_{H_c} - (M_{\eta_h} + M_{\eta_c})/2$, plotted
as a function of $M_{\eta_h}$. The open symbols correspond to values
extracted from lattice two-point functions, and the gray band 
is the fit to this data. We have shifted the value taken from experiments
(black burst) to compare directly with the data from the 
lattice calculation, which is missing electromagnetic and gluon 
annhilation effects.
\label{fig:1S_prelim}}
\end{figure}

The ground state $H_c$ energies are resolved very precisely
from the two-point correlator data. This is shown
in Fig.~\ref{fig:1S_prelim}, where we plot the quantity 
$\Delta_{H_c,hh} = M_{H_c} - (M_{\eta_h} + M_{\eta_c})/2$ 
as a function of $M_{\eta_h}$ 
determined from the corresponding heavy-heavy correlator. Because of this
the determination from lattice data is limited by systematic effects
due to missing electromagnetism, and the $\eta_b$ and $\eta_c$ being
unable to annihilate to gluons in our calculation. Using estimates of these
effects from~\cite{McNeile:2012qf}, we shift the experimental result downward
to compare directly with the lattice data.
With these taken into account, the lattice data is compatible with the
experimental value.
For comparison, the earlier lattice determination~\cite{McNeile:2012qf} 
using the same technique
but on $n_f = 2+1$ ensembles is also shown. In that case an error
was also estimated for the missing charm quarks in the sea, which are
included in the present study.

\subsection{$B_c(2S)$ energy}
\begin{figure}
\includegraphics[width=\textwidth]{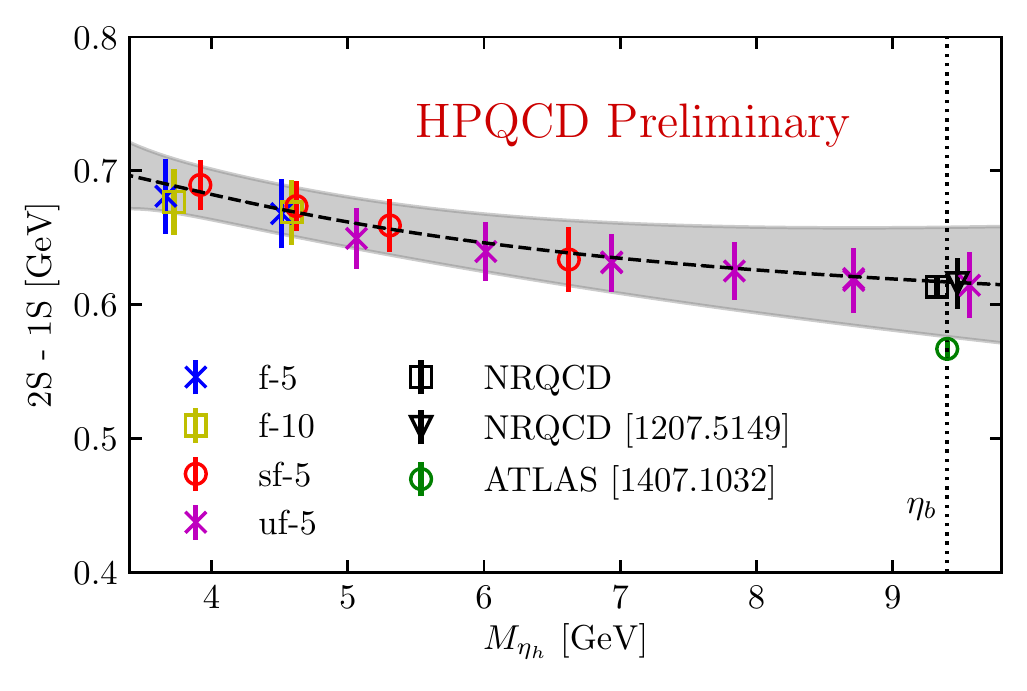}
\caption{Preliminary results for the $H_c$ 2S-1S splitting determined
from lattice two-point functions, as a function of $M_{\eta_h}$.
The fit to the lattice results is given by the gray band, 
with the physical result for the $B_c$ on the right at $\eta_b$, 
alongside the observation from 
ATLAS (green circle) and NRQCD results (black square and triangle). 
\label{fig:Bc_2S-1S}}
\end{figure}

Fig.~\ref{fig:Bc_2S-1S} shows our results for the $H_c$ 2S-1S splitting,
as a function of $M_{\eta_h}$. To fit the two-point correlators~\eqref{exp},
we provide Bayesian priors for the amplitudes $a_i$ of 0(1), and priors
for the ground state energy and excited state 
energy splittings, $M_i - M_{i-1}$. 
We have taken the priors for the mass splittings of $\approx 600$ MeV,
with a width that is $50 \%$ of the splitting (using $25 \%$ gives consistent
results, but with slightly less conservative errors). 
This covers the expected
range of the $2S$-$1S$ splitting from charmonium to bottomonium, 
and is consistent
with expectations for higher radial excitations~\cite{Ebert:2011jc}.
As shown in Fig.~\ref{fig:Bc_2S-1S}, the fit resolves the first
excited state energy within a typical
uncertainty of around 30 MeV. 
It should be stressed that only local correlators have been used in this
analysis, and that the inclusion of smearing functions should improve
the robustness of these determinations.

\begin{figure}
\includegraphics[width=\textwidth]{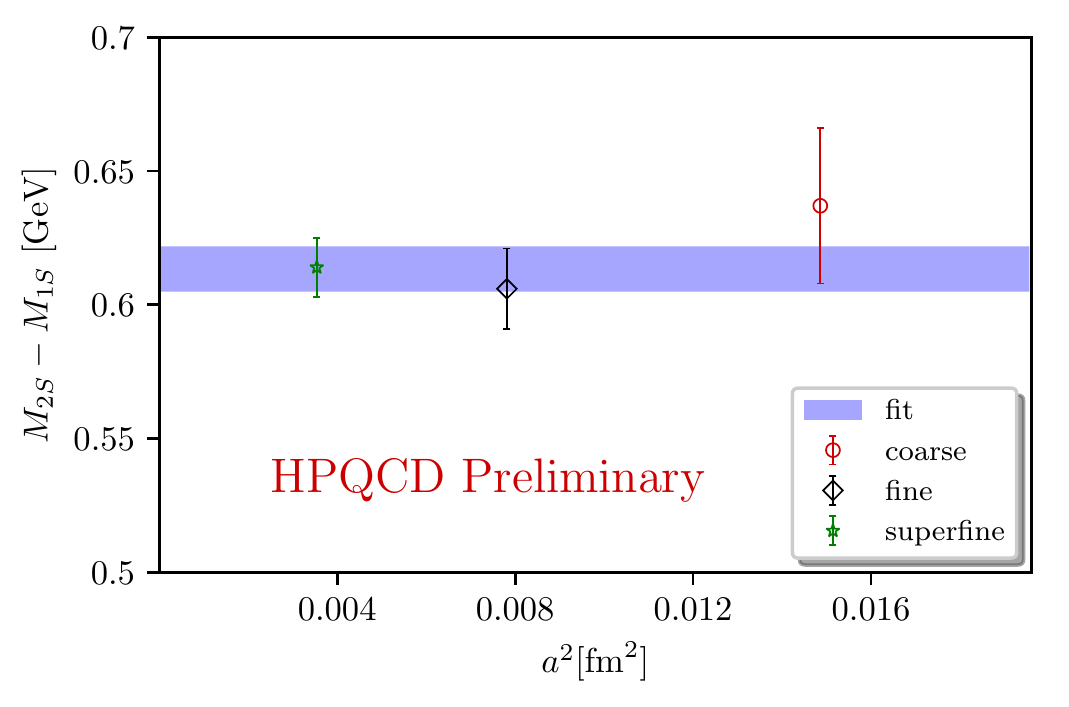}
\caption{The $B_c$ 2S-1S splitting obtained from $\O(\alpha_s)$-improved
 NRQCD $b$ quarks and HISQ charm quarks, calculated on the c-5 (red circle), 
 f-5 (black diamond), and sf-5 (green star) ensembles. The fit
 value corresponding to the $a=0$ determination is shown by the blue band.
\label{fig:Bc_splitting_fit_NRQCD}}
\end{figure}

As in the case of the ground state energy, there is no evidence
of significant discretisation artifacts in the data. 
We have included the data point from
the `uf-5' ensemble at $am_h = 0.9$ on the figure, but it is not used
in the fit given by the gray band. For comparison we also show
two NRQCD determinations of this quantity, 
from~\cite{Dowdall:2012ab} computed on $n_f = 2+1$ ensembles,
and from preliminary results computed here on the present ensembles and
shown in Fig.~\ref{fig:Bc_splitting_fit_NRQCD}.
These  agree well with the heavy-HISQ determination at $\eta_b$ and with
one another. The observation from ATLAS 
lies just outside of the one-sigma band from heavy-HISQ.

\section{Conclusions}
Here we have presented a preliminary result for the $B_c(2S)$ energy,
determined using fully relativistic heavy quarks, as well as NRQCD
$b$-quarks.
We find that we are consistenly 
able to resolve the $H_c$ radial excitation in our relativistic
two-point correlator data. 
The $2S$-$1S$ splitting falls slowly with $m_h$ over the range
from $m_c$ to $m_b$. There is no evidence of large discretisation
effects in the energy splitting, and for the moment we use a conservative
estimate of these effects.

With this methodology we find a preliminary value of 617(41) MeV for
the $2S$-$1S$ splitting, or using the PDG average for the 
$B_c$ mass 6274.9(8) MeV, 6892(41) MeV for the $2S$ energy.
This is consistent with NRQCD determinations of the same quantity
(\cite{Dowdall:2012ab} and Fig.~\ref{fig:Bc_splitting_fit_NRQCD}), 
but at present has larger error bars.
Both this result and the NRQCD results are above 
the ATLAS observation~\cite{Aad:2014laa},
at the level of one sigma for heavy-HISQ and two sigma for NRQCD.
Because the lattice QCD results have larger uncertainties than the ATLAS
result the significance of the discrepancy depends on the lattice errors.
One deficiency of the present calculation is that the HISQ
two-point correlators are computed 
without the use of smearing functions. 
Including one or more smearing functions
to expand the basis of correlators will help
resolve excited states in the lattice data, and is work in progress.

\section{Acknowledgements}
AL would like to thank C.\ Hughes and M.\ Padmanath for useful
discussions. 
This work was supported by the UK Science and Technology Facilities Council.
Computations were carried out on the DiRAC
Data Analytic system at the University of Cambridge,
operated by the University of Cambridge High Performance Computing Service, 
on behalf of the STFC DiRAC HPC facility. This is funded by BIS National
E-infrastructure and STFC capital grants and STFC DiRAC operation grants.

\end{document}